\documentstyle[twoside,fleqn,espcrc2]{article}

\def\greaterthansquiggle{\raise.3ex\hbox{$>$\kern-
.75em\lower1ex\hbox{$\sim$}}}
\def\lessthansquiggle{\raise.3ex\hbox{$<$\kern-
.75em\lower1ex\hbox{$\sim$}}}
\newcommand{\beq}{\begin{equation}}
\newcommand{\eeq}{\end{equation}}
\newcommand{\beqa}{\begin{eqnarray}}
\newcommand{\eeqa}{\end{eqnarray}}
\newcommand{\beqan}{\begin{eqnarray*}}
\newcommand{\eeqan}{\end{eqnarray*}}
\newcommand{\ba}{\begin{array}}
\newcommand{\ea}{\end{array}}
\newcommand{\no}{\nonumber}

\newcommand{\ra}{\rightarrow}

\newcommand{\wt}{\widetilde}

\newcommand{\A}{{\cal A}}

\newcommand{\F}{{\cal F}}
\newcommand{\G}{{\cal G}}

\newcommand{\M}{{\cal M}}

\newcommand{\U}{{\cal U}}

\newcommand{\st}{\stackrel}
\def\nz{\ifmmode {I\hskip -3pt N} \else {\hbox {$I\hskip -3pt N$}}\fi}
\def\zz{\ifmmode {Z\hskip -4.8pt Z} \else
       {\hbox {$Z\hskip -4.8pt Z$}}\fi}
\def\qz{\ifmmode {Q\hskip -5.0pt\vrule height6.0pt depth 0pt
       \hskip 6pt} \else {\hbox
       {$Q\hskip -5.0pt\vrule height6.0pt depth 0pt\hskip 6pt$}}\fi}
\def\rz{\ifmmode {I\hskip -3pt R} \else {\hbox {$I\hskip -3pt R$}}\fi}
\def\cz{\ifmmode {C\hskip -4.8pt\vrule height5.8pt\hskip 6.3pt} \else
       {\hbox {$C\hskip -4.8pt\vrule height5.8pt\hskip 6.3pt$}}\fi}

\def\au{{\setbox0=\hbox{\lower1.36775ex%
\hbox{''}\kern-.05em}\dp0=.36775ex\hskip0pt\box0}}
\def\ao{{}\kern-.10em\hbox{``}}
\def\lint{\int\limits}

\title{Quantizing Yang--Mills Theory: From Parisi-Wu Stochastic
Quantization 
to a Global Path Integral}

\author{Helmuth H{\"u}ffel\thanks{ Talk given at the  {\it Third Meeting on
Constrained Dynamics and Quantum Gravity}, Villasimius, Sardinia, Italy,
Sept. 
13-17, 1999; Univ. of Vienna preprint UWThPh-1999-62}
 and Gerald Kelnhofer\address{
       Institut f{\"u}r Theoretische Physik,
        Universit\"at Wien, Boltzmanngasse 5, A-1090 Wien}%
		\thanks{Supported by "Fonds zur F{\"o}rderung der wissenschaft\-lichen 
Forschung in \"Oster\-reich",  project P10509-NAW}}

\begin{document}

\begin{abstract}

Based on a generalization of the stochastic quantization scheme we 
recently proposed a generalized, globally defined 
Faddeev-Popov path integral density for the 
quantization of Yang-Mills theory. 
In this talk first our approach on the whole space of gauge potentials
is shortly reviewed; in the following we discuss the corresponding
global path 
integral on the gauge orbit space relating it to the 
original Parisi-Wu stochastic quantization scheme.

\vspace{1pc}
\end{abstract}

\maketitle

\section{MATHEMATICAL SETTING}

It is  our aim to discuss a globally valid  path integral procedure for
the 
quantization of Yang--Mills theory
based on a recently introduced
generalization \cite{annalsym,Vancouver,global,aspects}  of the Parisi--Wu
stochastic quantization scheme \cite{Parisi+Wu,Damgaard+Huffel,Namiki};
for 
different globally valid stochastic interpretations of the
Faddeev--Popov procedure 
\cite{Fadd} see \cite{asorey,Paycha}.

Let $P(M,G)$ be a principal fiber 
bundle with   compact structure group $G$ over
the compact Euclidean space time $M$. Let $\A$ denote the space of all
irreducible connections  on $P$ and let $\G$ denote the gauge group, 
which is given by all vertical automorphisms on $P$ reduced by the centre
of $G$. Then $\G$ acts freely on $\A$ and defines a principal 
$\G$-fibration
$\A \st{\pi}{\longrightarrow} \A/\G =: \M$ over the paracompact 
\cite{Mitter} space
$\M$ of all inequivalent gauge potentials
with projection $\pi$. 
Due to the Gribov ambiguity  \cite{Gribov} the principal 
$\G$-bundle $\A \ra \M$
is not globally trivializable. 
 
From \cite{Mitter} it follows that there exists a locally finite open 
cover $\U
=\lbrace U_{\alpha} \rbrace$  of $\M$ together with a set of background 
gauge fields $\lbrace A_{0}^{(\alpha) } \in \A
 \rbrace$  such that 
\beq
\Gamma_{\alpha} = \{ B \in 
\pi^{-1} (U_{\alpha})|D^{*}_{A_{0}^{(\alpha)}} (B - 
A_{0}^{(\alpha)}) = 0\}
\eeq
defines a family of local sections of $\A \ra \M$. Here 
$D_{A_{0}^{(\alpha)}}^*$ 
is the
adjoint operator of the covariant derivative $D_{A_{0}^{(\alpha)}}$ with 
respect
to $A_{0}^{(\alpha)}$.

\section{PARISI--WU STOCHASTIC QUANTIZATION}

The Parisi--Wu approach for the stochastic quantization
of the Yang--Mills theory is defined in terms of the Langevin equation
\beq
dA = - \frac{\delta S}{\delta A} ds + dW.
\eeq
Here $S$ denotes the Yang--Mills action without gauge symmetry breaking
terms and without accompanying ghost field terms,
$s$ denotes the extra time coordinate 
with respect to which the stochastic process is evolving, $dW$ is the
increment of a Wiener process. 

Instead of analyzing Yang-Mills theory in the original field 
space $\A$  we 
consider the family of trivial  principal $\G$-bundles 
$\Gamma_{\alpha} \times \G \ra \Gamma_{\alpha}$, 
which are locally isomorphic to the bundle 
$\A \ra \M$, where the isomorphisms are provided by the maps
\beq
\chi_{\alpha} : \Gamma _{\alpha} \times \G \ra 
\pi^{-1}(U_{\alpha}), \qquad
\chi_{\alpha} (B,g) := B^g
\eeq
with $B \in \Gamma_{\alpha}$, $g \in \G$ and 
$B^g$ denoting the  gauge 
transformation of $B$ by $g$. 

We transform the Parisi--Wu Langevin equation (3) into 
the adapted coordinates $\Psi = \left( \ba{c} B \\ g \ea \right)$. As 
this transformation is not globally possible the region of 
definition of (3) has to be restricted to
$\pi^{-1}(U(A_{0}^{(\alpha)}))$. Making use of the Ito stochastic calculus
 \cite{belo} the above Langevin equation now reads 
\beqa
\lefteqn{d\Psi =
 \left[-  G_{\alpha}^{-1} \frac{\delta S_{\alpha}
}{\delta \Psi}
+ \right. } \no \\
&& \left. \frac{1}{\sqrt{\det G_{\alpha}}} 
\frac{\delta( G_{\alpha}^{-1} \sqrt{\det G_{\alpha}
})}{\delta \Psi} \right] ds 
+  E_{\alpha} dW.
\eeqa
where $S_{\alpha}=\chi_{\alpha}^* S$ denotes the gauge invariant 
Yang-Mills action expressed in terms of the adapted coordinate $B$. 
The  
explicit form of the vielbein $E_{\alpha}$ corresponding to the change 
of coordinates $A \ra (B,g)$, the induced metric $G_{\alpha}$, 
its inverse and its determinant
can be found in \cite{annalsym}; for completeness we just recall that
\beq
\det G_{\alpha} = \nu^2 \, (\det \F_{\alpha})^2 \,
(\det \Delta_{A_{0}^{(\alpha)}})^{-1}.
\eeq
Here $\nu = \sqrt{\det (R_g^* R_g)}$ implies an invariant volume density 
on $\G$, 
where $R_g$  is the
differential of right multiplication
transporting any tangent vector in $T_g \G$ back to the identity 
$id_{\G}$  on $\G$ ; 
$\F_{\alpha} = D_{A_{0}^{(\alpha)}}^* D_B$
is the Faddeev--Popov operator.

\section{GENERALIZED STOCHASTIC QUAN\-TIZATION}

It is the basic idea of the  stochastic quantization scheme to study in 
addition to a given Langevin equation the associated Fokker--Planck
equation 
for the probability density $\rho$ interpreting
the equilibrium limit of 
this density as  Euclidean path integral measure. It is well 
known that such a procedure breaks down in the case of gauge theories,
as 
due to the gauge invariance no 
normalizable equilibrium limit emerges. A generalization of the
Parisi--Wu 
scheme was introduced in \cite{Zwanziger81} and extended 
recently  
in \cite{annalsym} by 
performing special,     
geometrically distinguished 
modifications  of both the drift and diffusion terms  such 
that -most 
essentially- all
 expectation values of gauge invariant observables are left unchanged. This
lead to  
a well defined Fokker--Planck formulation and the 
equilibrium density was derivable straightforwardly.
The Langevin equation (2) expressed in the adapted coordinates $\Psi$
thus  gets recast into
\beqa
\lefteqn{d\Psi =
 \left[- \wt G_{\alpha}^{-1} \frac{\delta S_{\alpha}^{\rm tot}
}{\delta \Psi}
+ \right. } \no \\*
&& + \left. \frac{1}{\sqrt{\det G_{\alpha}}} 
\frac{\delta(\wt G_{\alpha}^{-1} \sqrt{\det G_{\alpha}
})}{\delta \Psi} \right] ds 
+ \wt E_{\alpha} dW.
\eeqa
Here $\wt E_{\alpha}$ and  
$\wt G_{\alpha}^{-1}  = \wt E_{\alpha} \wt E_{\alpha}^*$ 
denote a specific vielbein and a  (inverse) 
metric, respectively, which are associated  to  
the above mentioned modifications of the drift and diffusion term of (2) 
or (4), respectively. The geometric interpretation of these 
modifications 
was disucssed in full length in \cite{annalsym} and \cite{aspects} and 
will not be repeated here.
 $S_{\alpha}^{\rm tot}$
denotes a total Yang-Mills action  
\beq
S_{\alpha}^{\rm tot} = \chi_{\alpha}^* S + pr_{\G}^* S_{\G}
\eeq
defined by the original Yang-Mills action $S$
 and by $S_{\G} \in C^{\infty}(\G)$ 
which is an 
arbitrary functional on $\G$ such that $e^{-S_{\G}}$ is integrable 
with respect to  the invariant volume density
 $\nu$,  
$pr_{\G}$ is the projector $\Gamma_{\alpha} \times \G \ra \G$.

The Fokker--Planck equation associated to (6) can easily be deduced 
\beq
\frac{\partial \rho[\Psi,s]}{\partial s} = L[\Psi] \rho[\Psi,s],
\eeq
where the Fokker-Planck operator $L[\Psi]$ is 
appearing in just factorized form
\beqa
\lefteqn{L[\Psi] = 
 \frac{\delta}{\delta \Psi}
\,\, \wt G_{\alpha}^{-1} } \no \\ 
&& \left[ \frac{\delta S_{\alpha}^{\rm tot}}{\delta \Psi}
- \frac{1}{\sqrt{\det G_{\alpha}}} 
\frac{\delta(\sqrt{\det G_{\alpha}})}{\delta \Psi}
+ \frac{\delta}{\delta \Psi}
 \right].
\eeqa
Due to the positivity of $\wt G_{\alpha}$ the fluctuation dissipation
theorem 
applies and the  equilibrium  distribution (for a proper normalization 
condition see the next section)  is 
obtained  by mere inspection as
\beq
\mu_{\alpha} \, e^{-S_{\alpha}^{\rm tot}}, \quad 
\mu_{\alpha} = \sqrt{\det G_{\alpha}}.
\eeq

Although our result implies 
unconventional 
$finite$ contributions along the gauge group (arising from the 
$pr_{\G}^* S_{\G}$ term) it is equivalent to the usual 
Faddeev--Popov prescription \cite{Fadd} for Yang--Mills theory. This
follows from 
the fact that for expectation 
values of gauge invariant observables these contributions along the
gauge 
group are exactly canceled out due to the normalization of the path integral.
 We stress once more that 
due to the Gribov ambiguity the usual Faddeev--Popov 
approach as well as -presently- our modified version 
are valid only locally in field space. 

\section{GLOBAL PATH INTEGRAL}

In order to compare  expectation values on different patches we 
consider the  diffeomorphism $\phi_{\alpha_1 \alpha_2}$ in the 
overlap of the two patches 
$(\Gamma_{\alpha_1} \cap \pi^{-1}(U_{\alpha_2})) \times \G$ and 
$(\Gamma_{\alpha_2} \cap \pi^{-1}(U_{\alpha_1})) \times \G $ 
which is given by
 \beq
\phi_{\alpha_1 \alpha_2} (B,g) := 
(B^{\omega_{\alpha_2}(B)^{-1}}, g).
\eeq
Here the field dependent gauge transformation 
$\omega_{\alpha_2} : \pi ^{-1}(U_{\alpha_2}) \ra \G$ is 
uniquely defined   
by $A^{\omega_{\alpha_2}(A)^{-1}} \in \Gamma_{\alpha_2}$. 
To the  density  $\mu_{\alpha}$ there is associated  a corresponding 
twisted top  form on $\Gamma_{\alpha} \times \G$ 
 which for simplicity we denote by the same symbol.
Using for convenience a matrix 
representation of $G_{\alpha}$ \cite{annalsym} we 
straightforwardly verify that
\beq
\phi_{\alpha_1 \alpha_2}^* \, \mu_{\alpha_2} = \mu_{\alpha_1} \, .
\eeq
This immediately implies  that in overlap regions the 
 expectation values of gauge invariant observables $f \in
C^{\infty}(\A)$ 
 are equal 
when evaluated in different  patches.
Let $\gamma_{\alpha}$ be a partition of unity of $\M$. We
 propose the definition of the global expectation value of a 
gauge 
invariant observable $f \in C^{\infty}(\A)$ by summing over all  
$\gamma_{\alpha}$ such that 
\beq
\langle f \rangle = 
\frac{\sum_{\alpha} \lint_{\Gamma_{\alpha} 
\times \G} \mu_{\alpha} \, e^{-S_{\alpha}^{\rm tot}}
\chi_{\alpha}^* (f \pi ^* \gamma_{\alpha})}
{\sum_{\alpha} \lint_{\Gamma_{\alpha} \times \G} 
\mu_{\alpha} \, e^{-S_{\alpha}^{\rm tot}}
\chi_{\alpha}^*  \pi ^* \gamma_{\alpha} }.
\eeq
Due to (12) it is trivial to prove that the global 
expectation value $\langle f 
\rangle$ is independent of  the specific 
choice of the locally finite cover $\lbrace U_{\alpha} \rbrace$, 
of the  choice of the background 
gauge fields $\lbrace A_{0}^{(\alpha)} \rbrace$ 
and of the choice of the partition of unity $\gamma_{\alpha}$, 
respectively.

As already indicated in \cite{annalsym} these structures can equally be 
translated into the original field space $\A$. With the help 
of the partition of unity the locally 
defined densities $\mu_{\alpha}$ as well as $e^{-S_{\alpha}^{\rm tot}}$
can be pieced together to give a globally well defined 
twisted top form $\Omega$ on $\A$
\beqa
\lefteqn{\Omega = \sum_{\alpha}  \chi_{\alpha}^{-1 \, *}
(\mu_{\alpha} \,  e^{-S_{\alpha}^{\rm tot}}) \, \pi ^* \gamma_{\alpha} } 
\no \\
&=& \sum_{\alpha} \mu \, e^{-S -\chi_{\alpha}^{-1 \, *} pr_{\G}^* S_{\G}
} \, \,
\pi ^* \gamma_{\alpha}.
\eeqa
The second equation follows  as a further consequence of (12) and $\mu$ 
denotes the flat measure  on $\A$. We remark 
the absence of  functional
determinants in the path integral measure and point out  the additional 
unconventional
interactions  implied by the $\chi_{\alpha}^{-1 \, *} pr_{\G}^*S_{\G}$ 
terms.
The  global expectation value (13) then reads
\beq
\langle f \rangle = 
\frac{ \int_{\A} \Omega \, f} 
{\int_{\A} \Omega }
\eeq
which due to the discussion from above is independent of all
the particular local choices.

\section{GAUGE ORBIT SPACE FORMULATION}

In addition   to the global expressions (13) and (15)
 for the path integral in the whole 
 space of connections the generalized
stochastic quantization scheme also offers the 
possibility of deriving the corresponding  formulation on the gauge 
orbit space $\M$: We consider  the projections of either 
the original Parisi--Wu Langevin equation (4)  or of  the modified 
equation (6) onto 
the gauge invariant subspaces $\Gamma_{\alpha}$ described by the 
coordinate $B$; in both cases we obtain 
\beqa
\lefteqn{dB = \left[ -(G_{\alpha}^{-1})^{\Gamma_{\alpha}
\Gamma_{\alpha}} 
\frac{\delta S}
{\delta B} + \right. } \no \\
&& \left.\frac{1}{\sqrt{\det G_{\alpha}}} \frac{\delta((G_{\alpha}^{-1})^
{\Gamma_{\alpha} 
\Gamma_{\alpha}}\sqrt{\det G_{\alpha}})}
{\delta B} \right]ds \no \\
&& \mbox{} + E_{\alpha}^{\Gamma_{\alpha}} dW.
\eeqa
Notice that in local coordinates $(G_{\alpha}^{-1})^{\Gamma_{\alpha} 
\Gamma_{\alpha}}$ is
the pullback of the restriction on $U_{\alpha}$ of the inverse of a globally
defined metric on the gauge orbit space $\M$  induced by the natural
metric 
on $\A$; similarly $E_{\alpha}^{\Gamma_{\alpha}}$ is defined. 
Since
the locally defined equations (16) are transforming covariantly under
the local
diffeomorphisms issued by the coordinate transformations, using
\cite{belo} it is
straightforward to check that their further projections onto $\M$
are yielding a globally defined stochastic process.

In direct analogy to our  
derivation of the local Fokker--Planck densities (8) we obtain   that
the 
 Fokker--Planck equation associated to the projected 
Langevin equations (16) 
  has an equilibrium distribution  given 
 by just the gauge invariant part of the densities  (8)
\beq
\det \F_{\alpha} \,
(\det \Delta_{A_{0}^{(\alpha)}})^{-1/2} \,
e^{-\chi_{\alpha}^* S}.
\eeq
By using (12) we can prove explicitly that their projections onto $\M$
on overlapping 
sets of $\U$ agree giving rise to a globally well defined top 
form  $\tilde{\Omega}$ on 
$\M$. Furthermore we can show that the above expectation values (13) 
and (15) of 
gauge invariant observables $f$ can identically be rewritten as 
corresponding integrals  over the gauge orbit space $\M$ with respect to 
 $\tilde{\Omega}$ 
\beq
\langle f \rangle = 
\frac{ \int_{\M} \tilde{\Omega} \, f} 
{\int_{\M} \tilde{\Omega} }.
\eeq
We note that this last expression shows agreement with the formulation 
proposed by Stora \cite{Stora} upon identification of $\tilde{\Omega}$
with the Ruelle-Sullivan form. 
Whereas in \cite{Stora}, however,  this form of the
expectation values on 
$\M$ appeared as the starting point
for a global formulation of Yang-Mills theory in the whole space of 
gauge potentials it appears in our case as the final result; 
we  aimed 
at its direct
derivation  within the generalized stochastic quantization approach.

We see that   the projections onto the local gauge 
fixing surfaces $\Gamma_{\alpha}$ of in specific the
original Parisi--Wu stochastic process  induce a globally defined
stochastic process on the gauge orbit space yielding the construction of the
globally defined  path integral density $\tilde{\Omega}$. 
We conclude that the globally defined 
Parisi--Wu Langevin equation  (2) on the whole field space $\A$ is
intimately related 
to the globally
defined path integral density (18) on the gauge orbit space $\M$.

\end{document}